\documentstyle[aps,epsfig]{revtex}
\newcommand{\be}{\begin{equation}}
\newcommand{\ee}{\end{equation}}
\newcommand{\bea}{\begin{eqnarray}}
\newcommand{\eea}{\end{eqnarray}}

\newcommand \la{\raisebox{-.5ex}{$\stackrel{<}{\sim}$}}

\begin{document}

{\Large
\begin{center}
\bf{ Freeze-Out Time in Ultrarelativistic Heavy Ion Collisions
from Coulomb Effects in Transverse Pion Spectra}
\end{center} }

\begin{center}
H.W. Barz$^1$, J.P. Bondorf$^2$, J.J. Gaardh{\o}je$^2$, 
and H. Heiselberg$^3$\\[5mm]
$^1$
Institut f\"ur Kern- und Hadronenphysik, FZ Rossendorf, 
Pf 510119,\\ 01314 Dresden, Germany\\
$^2$Niels Bohr Institute, Blegdamsvej 17, 2100 Copenhagen, Denmark\\ 
$^3$NORDITA, Blegdamsvej 17, 2100 Copenhagen, Denmark

\vspace*{10mm}
\end{center}
\vspace*{10mm}
{\bf Abstract:}\\
The influence of the nuclear Coulomb field on 
transverse spectra of $\pi^+$ and $\pi^-$ measured in $Pb+Pb$ 
reactions at 158 A GeV has been investigated.
Pion trajectories are calculated in the field
of an expanding fireball.
The observed enhancement  of the $\pi^-/\pi^+$ ratio 
at small momenta depends on the temperature and transverse 
expansion velocity of the source,
the rapidity distribution of the  net positive charge, and mainly  
the time of the freeze-out.

\vspace*{5mm}


\vspace*{10mm}

\noindent
\newpage

In heavy ion collisions at ultrarelativistic energies, nuclear systems
with a total net positive charge of around $Z=160$ can now be created
and studied. It has recently been observed \cite{NA44} in central
Pb+Pb collisions at CERN-SPS energies (158 A GeV) that the ratio of
negative pions to positive pions is significantly increased at low
transverse masses ($m_\perp \la m_\pi + 50 MeV$) for central rapidities.
A possible cause for this effect is the Coulomb interaction between
the produced pions and the positive charge from the reaction
partners. Such effects are already known in reactions at intermediate
energies \cite{early}.

If the two incoming ions pass through each other without stopping, the
combined electromagnetic field of the two moving ions has only a
minimal effect on the motion of the pions that are produced around
midrapidity. On the average, a charged pion experiences at time
$\tau_f$ and transverse distance $R$ a momentum change of $\sim
Ze^2R/(\gamma_{cm}\tau_f)^2$, where $Z$ is the combined charge of the
target and projectile nuclei and $\gamma_{cm}$ is the Lorentz factor.
This effect is inversely proportional to the bombarding energy. On the
other hand, if the colliding ions are strongly decelerated (large
nuclear stopping) the total charge stays together for a sufficient
amount of time to accelerate or decelerate the produced charged pions
significantly. Under such conditions distortions of the pion spectra
can occur leading to non-uniform pion yield ratios.

In this letter we report on a study within a dynamical model of the
enhancement of the $\pi^-/\pi^+$ ratio as a function of the
properties of the participant fireball and investigate the sensitivity
to dynamical features such as the total participant charge, the
transverse expansion velocity and the temporal and spatial extent of
the source.

In relativistic heavy ion collisions at SPS energies most of the pions
are produced rather late in the collision history.  They are formed
either directly from strings or in the decay of resonances. In the
dense medium they undergo several collisions so that the pionic
ensemble comes close to thermal equilibrium, as can be deduced from
properties of measured transverse spectra. Due to the long range
nature of the Coulomb force the details of the production mechanism
are not very important for our purpose. We thus consider the system at
a proper time $\tau_f$ sufficiently large that the violent stage of
the collision has ceased, i.e. close to freeze-out.  Already at this
time a certain charge separation has occurred during the equilibration
phase due to the inhomogeneous Coulomb field.  However, the main
Coulomb acceleration effect occurs after freeze-out during the free
motion in the field of the source.  At this time the pions already
have a spectrum very similar to that measured in experiment, since
only the Coulomb field leads to further significant modification of
the spectrum shape, particularly for slowly moving particles.

We assume that the pions can be described by a thermal 
distribution (characterized by a temperature $T$) that 
is superimposed 
on a collective outward flow of the matter  
characterized by a 4-velocity field $u^\mu$
\begin{equation}
   u^\mu=(\gamma_\perp \cosh{y'},\gamma_\perp \sinh{y'},
{\bf u}_\perp), \quad \gamma_\perp = \sqrt{1+{\bf u}_\perp^2},\quad
{\bf u}_\perp(r) = \frac{\bar \beta}{\sqrt{1-{\bar \beta}^2}} 
                    \frac{{\bf r}_\perp}{R}.
\end{equation}
This field describes slices which move with rapidity
$y'$ in longitudinal z-direction (i.e. along the beam axis).
We assume that the reaction zone has an axially symmetric
cylindrical shape  and that the 
the transverse 4-velocity ${\bf u}_\perp$ scales 
linearly with the radial
distance ${\bf r}_\perp$ from the axis.
The transverse velocity $\bar \beta$
describes the motion of
a characteristic radius $R$.
 
Assuming that the rapidity slices follow a
Gaussian distribution centered at the center-of-mass rapidity 
$y_{cm}$ the pion source distribution at freeze-out reads
\begin{eqnarray} \nonumber
\frac{d^6N}{dyd{\bf p}_\perp^2 dzd{\bf r}_\perp^2}
\sim \int dy' \exp{[-\frac{(y'-y_{cm})^2}{2\Delta y^2}]}\,
 \delta(z-\tau_f \sinh{y'})F(\frac{r_\perp}{R})\\
\times \frac{u_\mu p^\mu}{\exp(u_\mu(p^\mu-eA^\mu)/T)-1}\, ,
\end{eqnarray}
where $p^\mu$, $e$,
and $A^\mu$ denote the pion momentum and charge and the 
electromagnetic 4-potential of the source, respectively. 
The distribution is a function of the rapidity $y$, the transverse
momentum ${\bf p}_\perp$, and the longitudinal and transverse
extensions $z$ and ${\bf r}_\perp$. In the spirit of the Bjorken
picture the longitudinal distance $z$ is related to the rapidity $y'$
of the cells via an effective freeze-out time $\tau_f$ although in
reality there will be a distribution of freeze-out times. The width
$\Delta y$ is taken from experiment.  For the transverse density
profile $F(r_\perp/R)$ we find that two different profiles of the
transverse pion distribution, a sharp cut-off at radius $F(r_\perp)
\sim\Theta(1-r_\perp /R)$ and a Gaussian distribution
$\exp{(-2r^2_\perp /R^2)}$ both with the same mean square radius,
give similar pion ratios within a few percent.
Due to the Coulomb potential $A^\mu$ the shapes of the distributions are
different for positively and negatively charged pions. 

The subsequent motion of the pions in the expanding electromagnetic
field created by the net charge is described by the 4-velocity $w^\mu$
which satisfies the equation of motion in the electromagnetic
potential $A^\mu$
\begin{equation}
m \frac{d}{d\tau} w^\mu= e w_\nu(A^{\nu,\mu}-A^{\mu,\nu}),
\end{equation}
where $\tau$ denotes the proper time and $m$ is the pion
mass. The final spectra are obtained 
by sampling over a set of different initial conditions using eq. (2).

The potential $A^\mu$ is obtained from the current distribution of the
(positive) net charge. Most of this charge is carried by the baryons
which are not fully stopped \cite{NA44prot,NA49}
and, thus, may have a wider rapidity
distribution than the pions which are preferentially produced at
midrapidity. 
In the case of symmetric reactions we
parametrize the charge distribution as
\begin{eqnarray}
\nonumber
f_{ch}(y) =&& Z_{cent}\frac{1}{\sqrt{8\pi}\Delta y_{ch}}
          [\exp{(-\frac{(y-y_1)^2}{2\Delta y_{ch}^2})}+
           \exp{(-\frac{(y+y_1)^2}{2\Delta y_{ch}^2})}]\\ 
         +&&z_t\delta(y-y_t) + z_p\delta(y-y_p),
\end{eqnarray}
describing the participant net charge $Z_{cent}$ in the fireball
as two distributions of width 
$\Delta y_{ch}$ centered at rapidities $\pm y_1$ in the center
of mass system. For non-central collisions the rest charges 
$z_{t,p}$ of the target and projectile continue to move with their 
original rapidities $y_t$ and $y_p$. 

The participant zone further expands due to the collective transverse
flow (1) and the random thermal motion. We model the time evolution of
the charge distribution by a Monte-Carlo sample of charged test
particles. These test particles move on straight trajectories with
initial conditions given by eq. (2) but now with the longitudinal
distribution (4) instead of the single Gaussian distribution. The
potential $A^\mu$ is then calculated by summing up the retarded
potentials of the test particles.  Due to the large retardation
effects an essential part of the electromagnetic potential is
generated by the charge distribution prior to freeze-out.  We describe
this situation between $0<\tau<\tau_0$ by the hydrodynamic flow (2) of
the matter.  During this period the transverse flow is assumed to
increase linearly with time.  The potential depends somewhat on the
particle composition of the charged zone since the thermal velocity of
the particles depends on their masses.  We have included a 10\% excess of
negative pions over positive pions compensated by the 
positive kaon net charge \cite{Jones}.

In Fig. 1a-c we compare the result of the calculations to recent data
from the NA44 experiment for $Pb+Pb$ collisions at $E=158 A GeV$ and
for $S+ Pb$ and $S+ S$ at $200 A GeVA$ \cite{NA44}, all for central
collisions and rapidities $3.3<y<4.0$. The calculations take the
actual NA44 detector acceptance \cite{NA44} into account.  In
addition, we also plot calculations corresponding to a detector
located exactly at midrapidity. Since absolute $\pi^-/\pi^+$ ratios
are not available, we normalize our calculated ratios to unity in the
region $200 MeV < m_t-m_\pi < 400 MeV$, as is done for the
experimental results.

Lead data were taken with a trigger selecting 15\% of the total
interaction cross section. In a sharp cut-off model this implies an
average charge of $Z_{cent}=122$ in the fireball. In the calculations
we fix the charge of the participant zone to this value. We transform the
transverse area into a circle of $R_{geom}=5.8$ fm.  At freeze-out
we take
a radius of $R=R(\tau_f)= R_{geom} + \bar \beta
\tau_f/2$ assuming  the linear increase of the transverse velocity.
The widths of the pion rapidity distributions was chosen to be 
$\Delta y =1.3$ 
corresponding to measurements in $S + S$,
$S+Pb$ \cite{S+S} and with measured transverse energy spectra in
$Pb+Pb$ collisions\cite{Alber}.
We use a temperature of  $T=120 MeV$ and  
$\bar \beta $= 0.62  corresponding to a mean transverse expansion
velocity $\langle \beta \rangle = 0.42$. 
These parameters are well compatible with
the values obtained from a systematic analysis of measured 
transverse $\pi, K$ and $p$ spectra from $Pb +Pb$ data 
\cite{NA44slopes}, where a definite anticorrelation between extracted
temperature and transverse velocity was established. 
Recent
$dN/dy$ vs $y$ data from the NA44 \cite{NA44prot} and
NA49\cite{NA49} experiments provide us with the rapidity distribution
of protons which can be characterized by  the parameters
$\Delta y_{ch}=0.84$  and  $y_1=1.1$, implying a
rapidity loss of 1.7.

Using these values a good description of the experimental data (see
Fig. 1a) was obtained using a freeze-out time of 7 fm/c.  Figs. 1b and
1c show a comparison for lighter systems.  Here we use a higher
temperature of 130 MeV and a smaller velocity $\langle \beta \rangle =
0.29$ \cite{S+S,braun}.  A fit to the $S$ + $S$ data \cite{S+S}
indicates a smaller rapidity loss of 1.3, leading to $\Delta
y_{ch}=1.25$, $y_1=1.7$.  For the asymmetric $S+Pb$ reaction a good
description of the data is obtained (see Fig. 1b) by assuming that 30
protons from the $Pb$ target and all protons from the projectile
participate in the fireball decelerated to their respective rapidities
$y_1$ in eq. (4).

The temperatures, transverse flow and charge distributions used in
Fig. 1 were obtained from best fits to experimental $p_\perp$ and
rapidity distributions.  The effect of varying the various parameters
is exhibited in Fig 2.  The magnitude of the enhancement clearly
scales with the total participant charge and diminishes if the charge
is distributed over a wider rapidity range (Fig. 2a). If the system
expands fast, the slower pions are overtaken by the expanding
potential and experience a smaller net charge.  This can be caused by
increasing collective flow (Fig. 2b) or by increasing thermal motion
(Fig. 2c) and results in a decrease of Coulomb effects.  A stronger
effect is observed when the freeze-out time $\tau_f$ increases, see
Fig. 2d, reflecting a more diluted charge distribution.  The profile
of the rapidity distribution of the net charge also plays a role.
Further, we remark that the pion ratio decreases slightly from 1.60 to
1.55 if the net charges of kaons and pions are not taken into account.

Hanbury-Brown and Twiss (HBT) analysis 
provides two independent comparisons of the freeze-out time.  In a
longitudinally expanding system with Bjorken scaling the freeze-out
time is related to the longitudinal HBT radius by $\tau_f = R_l\,
\sqrt{m_\perp/T}$. From the NA44 data $R_l\simeq 5.5$fm \cite{NA44HBT}
we estimate $\tau_f\simeq 8$fm/c. From transverse HBT radii we obtain
$R=2R_s\simeq 10$fm. Comparing to $R=R_{geom}+\bar{\beta}\tau_f/2$
we deduce $\tau_f=8$fm/c; this value should, however, be corrected for
transverse flow which affects the HBT radii. We conclude that both
freeze-out times extracted from the longitudinal and transverse HBT
radii are compatible with the one we find from $\pi^-/\pi^+$.
We also remark that RQMD calculations predict substantially 
larger values of $\tau_f$=15 fm/c \cite{Sorge}.

A simple estimate of the Coulomb effect can be made from the retarded
electric field resulting from a net charge distribution, $dN^{ch}/dy$,
that is approximately constant in rapidity.  Such a longitudinally
streaming charge distribution generates in transverse direction 
an electric field
${\bf E}\simeq 2e(dN^{ch}/dy) {\bf r}_\perp /(tR^2)$.
Here, $t$ is the time
after collision and $R$ is the transverse size of the charge distribution.
In average this field leads to a momentum change of the $\pi^\pm$ by
\begin{eqnarray}
  \frac{ \delta{\bf p}_\perp}{{\bf p}_\perp} \; \simeq \; 
  \pm e^2 \frac{dN^{ch}}{dy} \frac{2f}{mR}
  \,\; \equiv \; \pm \Delta
  \,,\label{pf}
\end{eqnarray}
for small transverse momenta.  The factor $f$ is of order the inverse
of the average proton velocity but also
depends on the freeze-out time and the transverse expansion
\cite{details}.  Assuming a
thermal transverse momentum distribution at freeze-out,
$dN/dp^2_\perp\sim\exp(-m_\perp/T)$, as has been used 
the spectra \cite{NA44slopes}, we obtain for small transverse momenta
after correcting for the Coulomb effect of Eq. (\ref{pf})
\begin{eqnarray}
 \frac{N(\pi^-)}{N(\pi^+)} =\frac{dN^-/dp_\perp^2}{dN^+/dp_\perp^2}
 \simeq 1 +4\Delta(1-\frac{m_\perp-m}{T}+ ...)\,. \label{ratio}
\end{eqnarray}
This approximate result shows that the linear decrease of the ratio
takes place on a scale of $m_\perp-m\simeq T$. The enhancement for
Pb+Pb collisions at midrapidities can be estimated using
$dN^{ch}/dy\simeq 35$ \cite{NA49}, $R\simeq10$ fm and $f\simeq 2$ to
be $4\Delta\simeq 0.6$.  This approximate value and
Eq. (\ref{ratio}) are in qualitative agreement with the pion ratio in
Fig. 1.

We have also calculated the kaon ratio $K^-/K^+$ in $Pb+Pb$
collisions using the same parameter set as used for the pions
(Fig. 1d).  As the kaon is heavier we expect
the corresponding $K^-/K^+$ ratio to be smaller by a factor of
$m_K/m_\pi \sim 3$ accordingly to Eq.(\ref{pf}).

In summary, we have found that the observed $\pi^-/\pi^+$ ratios can
be qualitatively explained as resulting from the Coulomb acceleration
of pions by the positive participant charge in a rapidly expanding
system.  We have demonstrated that a measurement of this charge ratio
in large systems supplements the standard HBT analyses providing an
independent constraint on the freeze-out time scale, the flow, and the
transverse size. The charge ratio is mostly influenced by the
freeze-out time. The quantitative agreement with experimental $Pb+Pb$
data requires a relatively short freeze-out time of about 7 fm/c
assuming a mean transverse flow velocity of $\langle\beta\rangle=0.42$
and a temperature of 120 MeV. Such values are compatible with analyses
of transverse pion, kaon and proton spectra. Larger freeze-out times
require less amount of flow.  The freeze-out time obtained is
compatible with freeze-out times and source sizes extracted in HBT
analyses.

Finally, we note that the
magnitude of the Coulomb effects in pion spectra are sensitive to the
degree of stopping and the resulting distribution of the positive
charge.  At RHIC and LHC energies the stopping and in particular the
net charge at midrapidity are expected to be smaller, and 
consequently we predict (see Eq.(6) and Fig. 2a) correspondingly 
smaller  Coulomb effects.

We acknowledge fruitful discussions with Dr. Nu  Xu (Los Alamos).
The support of the German BMBF and the Danish Natural Science 
Research Council is appreciated.

\newpage


\newpage

\begin{figure}
\centerline{
\psfig{figure=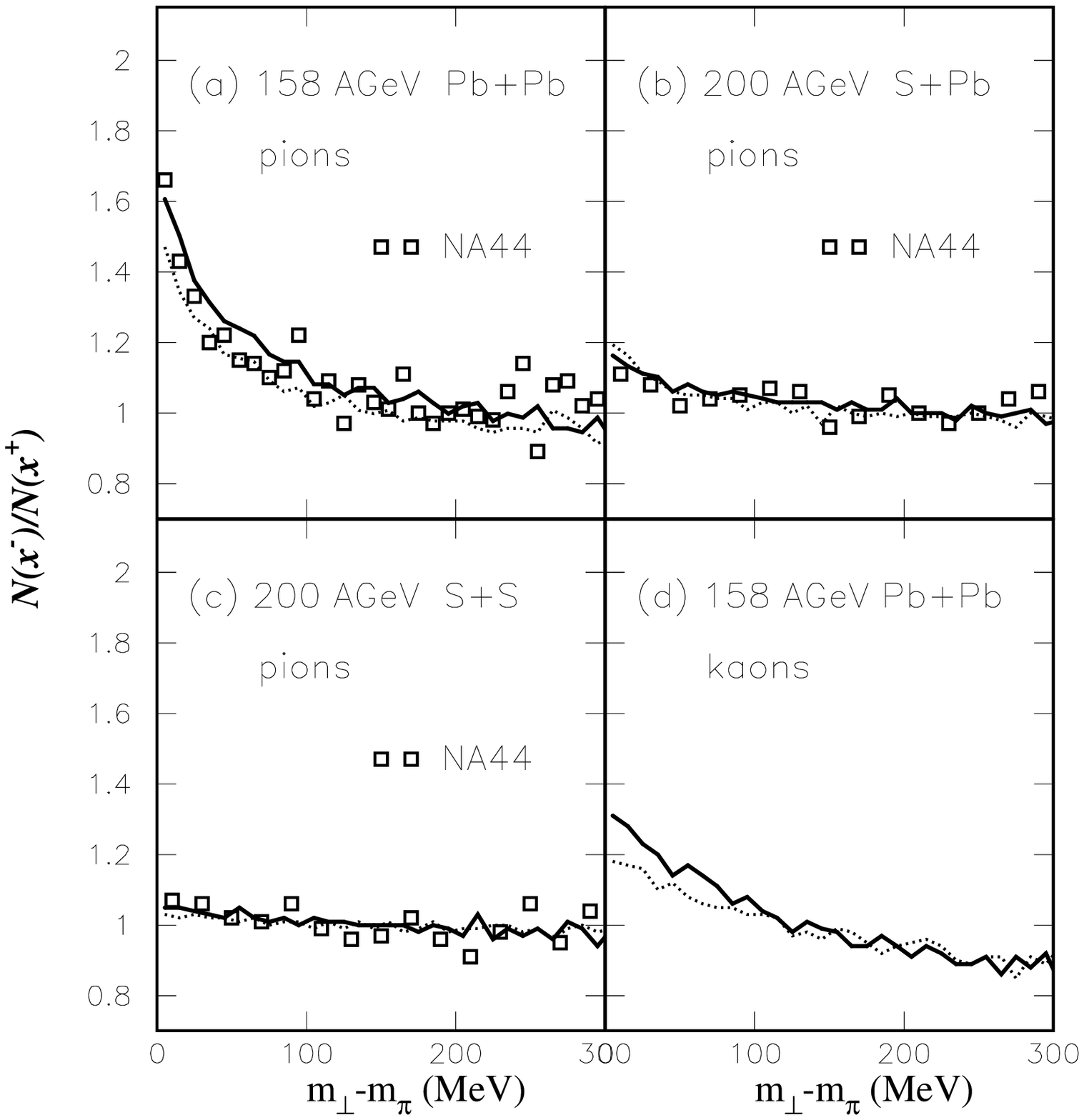,width=16cm,height=20cm,angle=0}}
\vspace{1cm}
\caption{
 Experimental $\pi^-/\pi^+$ ratios 
\protect\cite{NA44}  
for the reactions
(a) $Pb + Pb$, (b) $S + Pb$, and (c) $S + S$ 
as a function of transverse mass compared to  
calculations (solid lines) using the dynamical 
Coulomb model with a best-fit freeze-out time of 7 fm/c. 
The dotted lines show the results for a detector
located exactly at midrapidity.
Panel (d) shows the predicted arbitrarily normalized 
$K^-/K^+$ ratio for the $Pb + Pb$ reaction.
}
\end{figure}

\newpage

\begin{figure}
\centerline{
\psfig{figure=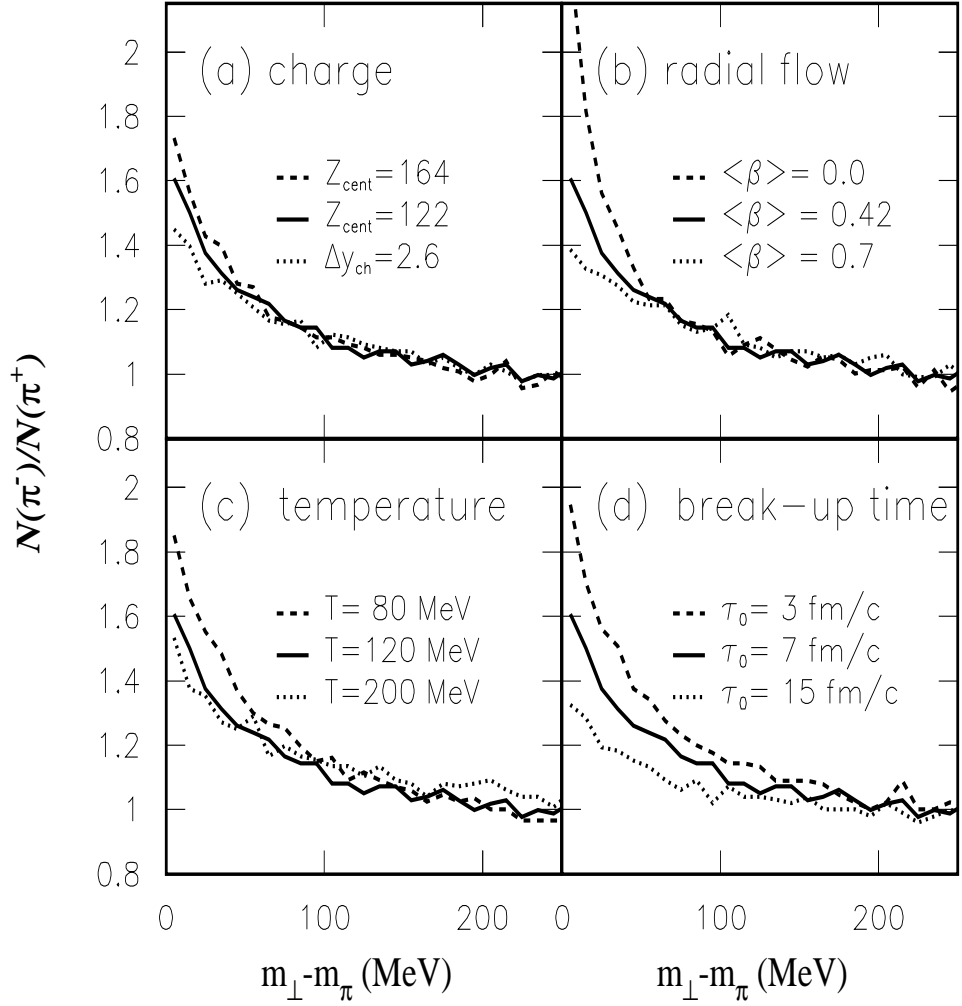,width=16cm,height=20cm,angle=0}}
\vspace{1cm}
\caption{
 Comparison of the sensitivity of the calculations to 
variation of the main  parameters. Solid lines represent 
the best fit calculations for the $Pb + Pb$ reaction shown in 
Fig. 1. Variation of : (a) total participant charge
$Z_{cent}$ and width of rapidity distribution, 
(b) average transverse expansion velocity $\langle \beta \rangle $,
(c) temperature at freeze-out, (d) freeze-out time $\tau_f$.
}
\end{figure}

\end{document}